\documentclass[runningheads]{llncs}
\usepackage{amsmath,amssymb}
\usepackage{graphicx}
\usepackage{todonotes,lineno}
\usepackage{paralist,soul}
\usepackage{rotating,multicol}
\usepackage{booktabs,tabularx}
\graphicspath{{figures/}}
\usepackage[font=small,labelfont=small]{subcaption}
\usepackage[font=small,labelfont=bf,format=plain]{caption}
\usepackage[pdfpagelabels,colorlinks,citecolor=blue,linkcolor=blue,urlcolor=blue]{hyperref}

\newcommand{\myparagraph}[1]{\medskip\noindent\textbf{#1}.}
\newcommand\rurl[1]{\href{http://#1}{\nolinkurl{#1}}}

\author{Michael~A.~Bekos\inst{1}, Mirco~Haug\inst{2}, Michael~Kaufmann\inst{2}, Julia~M\"annecke\inst{2}}

\authorrunning{M.~A.~Bekos, M.~Haug, M.~Kaufmann, J.~M\"annecke}
\title{An Online Framework to Interact and Efficiently Compute Linear Layouts of Graphs} 
\titlerunning{An Online Framework to Interact and Compute Efficiently Linear Layouts}

\institute{
Department of Mathematics, University of Ioannina, Ioannina, Greece\\
\texttt{bekos@uoi.gr}
\and
Institut f\"ur Informatik, Universit\"at T\"ubingen, T\"ubingen, Germany\\
\texttt{michael.kaufmann@uni-tuebingen.de}\\
\texttt{\{mirco.haug,julia.maennecke\}@student.uni-tuebingen.de}
}

\begin{document}
\maketitle

\begin{abstract}
We present a prototype online system to automate the procedure of computing different types of linear layouts of graphs under different user-specific constraints. Currently, four different types of linear layouts are supported: stack~\cite{DBLP:journals/jct/BernhartK79}, queue~\cite{DBLP:journals/siamcomp/HeathR92}, rique~\cite{DBLP:conf/gd/BekosFKKKR22} and deque~\cite{DBLP:conf/gd/AuerBBBG10}, as well as, any mixture of them. 

The system consists of two main components; the client and the server sides. The client side is built upon an easy-to-use editor, which supports basic interaction with graphs, enriched with several additional features to allow the user to define and further constraint the linear layout to be computed. 
The server side, which is available to multiple clients through a well-documented API, is responsible for the actual computation of the linear layout. Its algorithmic core is an extension of a SAT formulation~\cite{DBLP:conf/gd/Bekos0Z15}~that is known to be robust enough to solve non-trivial instances in reasonable amount of time. 
\end{abstract}

\section{Introduction}
\label{sec:introduction}

Linear layouts of graphs have been fruitful subjects of intense research over the years, both from a combinatorial and from an algorithmic point of view, as they play an important role in various fields; see, e.g.,~\cite{DBLP:journals/dmtcs/DujmovicW04}. Formally, a linear layout of graph $G=(V,E)$ consists of a linear order of its vertices (that is, a bijective function $\sigma: V \rightarrow \{1,\ldots,|V|\}$), and a partition of its edges into a particular number of sets. Different constraints on the edges that may reside in the same set, give rise to different types of linear layouts; see, e.g.,~\cite{DBLP:journals/tcs/AlamBDGKP21,DBLP:journals/ejc/BinucciGHL18,DBLP:journals/siamcomp/HeathR92,DBLP:conf/gd/Pupyrev17,DBLP:journals/jcss/Yannakakis89}. In that aspect, there is a rich body of research on specific types of linear layouts that are derived by leveraging different data structures to capture the order of the~vertices.

Formally, given $k$ data structures $\texttt{D}_1,\ldots,\texttt{D}_k$, a graph $G$ admits a $(\texttt{D}_1,\ldots,\texttt{D}_k)$-layout if there is a linear order $\prec$ of the vertices of $G$ and a partition of the~edges of $G$ into $k$ sets $\texttt{E}_1,\ldots,\texttt{E}_k$, called \emph{pages}, such that for each page $\texttt{E}_i$ in the partition, each edge $(u,v)$ of $\texttt{E}_i$ is processed by the data structure $\texttt{D}_i$ by inserting $(u,v)$ to $\texttt{D}_i$ at $u$ and removing it from $\texttt{D}_i$ at $v$ if $u \prec v$ in the linear layout. If $\texttt{D}_1,\ldots,\texttt{D}_k$ are not all of the same type, then the corresponding linear layout is called \emph{mixed}. Otherwise, the corresponding linear layout takes its name from the used data structure and one naturally seeks in minimizing the total number of needed pages. In this regard, the most studied types of linear layouts are the ones obtained by stacks, queues, restricted-input double-ended queues (or riques, for short) and general double-ended queues (or deques, for short). 

\begin{figure}[t]
	\centering
	\subcaptionbox{\label{fig:goldner-harary}}{\includegraphics[width=0.172\textwidth,page=1]{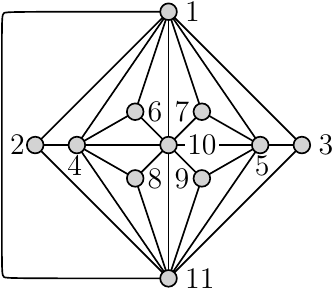}}
	\hfil
	\subcaptionbox{\label{fig:stack}}{\includegraphics[width=0.4\textwidth,page=2]{graphs}}
	\hfil
	\subcaptionbox{\label{fig:queue}}{\includegraphics[width=0.4\textwidth,page=3]{graphs}}
   \caption{%
   Illustration of:   
   (a)~the Goldner-Harary graph~\cite{GH75},  
   (b)~a $3$-stack layout of it, and
   (c)~a $2$-queue layout of it.}
\label{fig:sample}
\end{figure}

\myparagraph{Known Results} There exists a plethora of theoretical results for each of the aforementioned types of linear layouts; in the following, we overview existing results for planar graphs. For a more detailed overview, we point the interested reader to~\cite{DBLP:journals/dmtcs/DujmovicW04}. 
\begin{itemize}[--]
\item For stack layouts, the most notable result is to due Yannakakis, who back in 1986 showed that every planar graph admits a $4$-stack layout~\cite{DBLP:conf/stoc/Yannakakis86,DBLP:journals/jcss/Yannakakis89} improving a series of earlier results~\cite{DBLP:conf/stoc/BussS84,DBLP:conf/focs/Heath84,Istrail1988a}. Notably, the bound of $4$ was recently shown to be worst case optimal~\cite{DBLP:journals/jocg/KaufmannBKPRU20,DBLP:journals/jctb/Yannakakis20}. Note that several subfamilies of planar graphs allow for layouts with fewer than four stacks; see, e.g.,~\cite{DBLP:journals/algorithmica/BekosGR16,DBLP:journals/jct/BernhartK79,DBLP:journals/mp/CornuejolsNP83,DBLP:journals/dcg/FraysseixMP95,Ewald1973,DBLP:conf/focs/Heath84,DBLP:journals/appml/KainenO07,NC08,DBLP:conf/cocoon/RengarajanM95}. 
\item For queue layouts, a breakthrough result is due to Dujmovi\'c et al.~\cite{constant}, who showed that every planar graph admits a $49$-queue layout, improving previous (poly-)logarithmic bounds~\cite{DBLP:journals/corr/BannisterDDEW18,DBLP:journals/jgaa/DujmovicF18,DBLP:journals/siamcomp/BattistaFP13}. While this bound was slightly improved to $42$~\cite{DBLP:journals/algorithmica/BekosGR23}, the best-known corresponding lower bound is four~\cite{DBLP:journals/algorithmica/AlamBGKP20}, which implies that the current gap between the two bounds is still very large. Again, several subfamilies of planar graphs allow for layouts with significantly fewer than $42$ queues; see, e.g.,~\cite{DBLP:journals/algorithmica/AlamBGKP20,Ganley95,DBLP:journals/siamdm/HeathLR92,DBLP:journals/siamcomp/HeathR92,DBLP:conf/cocoon/RengarajanM95}.

\item For rique and deque layouts the literature is significantly reduced. Bekos et al.~\cite{DBLP:conf/gd/BekosFKKKR22} and Auer et al.~\cite{DBLP:conf/gd/AuerBBBG10} provide characterizations of the graphs admitting $1$-rique and $1$-deque layouts, respectively. While all planar graphs admit $2$-deque layouts~\cite{DBLP:conf/stoc/Yannakakis86}, the corresponding exact bound is not yet known for rique layouts (it ranges between $2$ and $4$).   
\end{itemize}
\myparagraph{Motivation} The primary motivation for the development of this system was a paper by Yannakakis, which had appeared at STOC in 1986~\cite{DBLP:conf/stoc/Yannakakis86} and contained a sketch of a proof for the existence of a planar graph that does not admit a $3$-stack layout. The details of this proof, however, had not appeared in a paper till very recently (in particular, the proof-sketch was not part of the subsequent journal version~\cite{DBLP:journals/jcss/Yannakakis89} of the extended abstract that had appeared at STOC~\cite{DBLP:conf/stoc/Yannakakis86}). Thus, the problem of determining whether there exists a planar graph that requires four stacks in any of its stack layouts was open for years, and clearly formed the most intriguing open problem in the field. 

Our effort to give an answer to this open problem led to the development of this system. Based on a previous work containing a SAT formulation~\cite{DBLP:conf/gd/Bekos0Z15} for the stack layout problem (that is, for the problem of finding a stack layout of a given graph with a certain number of stacks), we continued the work extending it with several new features that are of independent value, even for future considerations. Notably, using the system we managed to solve the aforementioned open problem in~\cite{DBLP:journals/jocg/KaufmannBKPRU20}, almost simultaneously with Yannakakis~\cite{DBLP:journals/jctb/Yannakakis20}, who also provided the details of his proof (almost three decades after the publication of the sketch). 

\myparagraph{Contribution} We describe a novel system, which automates the procedure of computing different types of linear layouts of graphs (i.e., stack-, queue-, rique-, deque-layouts or mixtures of these) by supporting standard procedures that a domain expert needs when interacting with such layouts. Besides the actual computation, the system is featured with an easy-to-use graph editor, which supports basic interaction with graphs, and simultaneously provides the necessary functionality to define and further constraint the linear layouts to be computed (e.g., by constraining the relative order of the vertices, or the edges to appear at specific pages e.t.c.). The system is available online at: 

\medskip\centerline{\url{http://alice.math.uoi.gr}}

\myparagraph{Paper Organization} In Section~\ref{sec:system}, we describe in details the extended features that we have implemented in the system together with some insights on the SAT formulation. Section~\ref{sec:proof-of-concept} serves as a proof of concept for our system, as we present several findings that have been derived using the system. We conclude in Section~\ref{sec:conclusions} with further considerations and plans.

\section{Description of the System}
\label{sec:system}

Our system, as introduced in Section~\ref{sec:introduction}, consists of two main components, the client and the server sides, and introduces a series of innovations over its previous implementation~\cite{bob}. The actual code of the system is available to the community at a \texttt{github} repository (\rurl{github.com/linear-layouts/SAT}). In the following, we describe in details the extended features that are currently supported both in the client (Section~\ref{subsec:client}) and the server side (Section~\ref{subsec:server}). 

\subsection{The Client Side}
\label{subsec:client}

\begin{figure}[t]
	\centering
	\includegraphics[width=\textwidth]{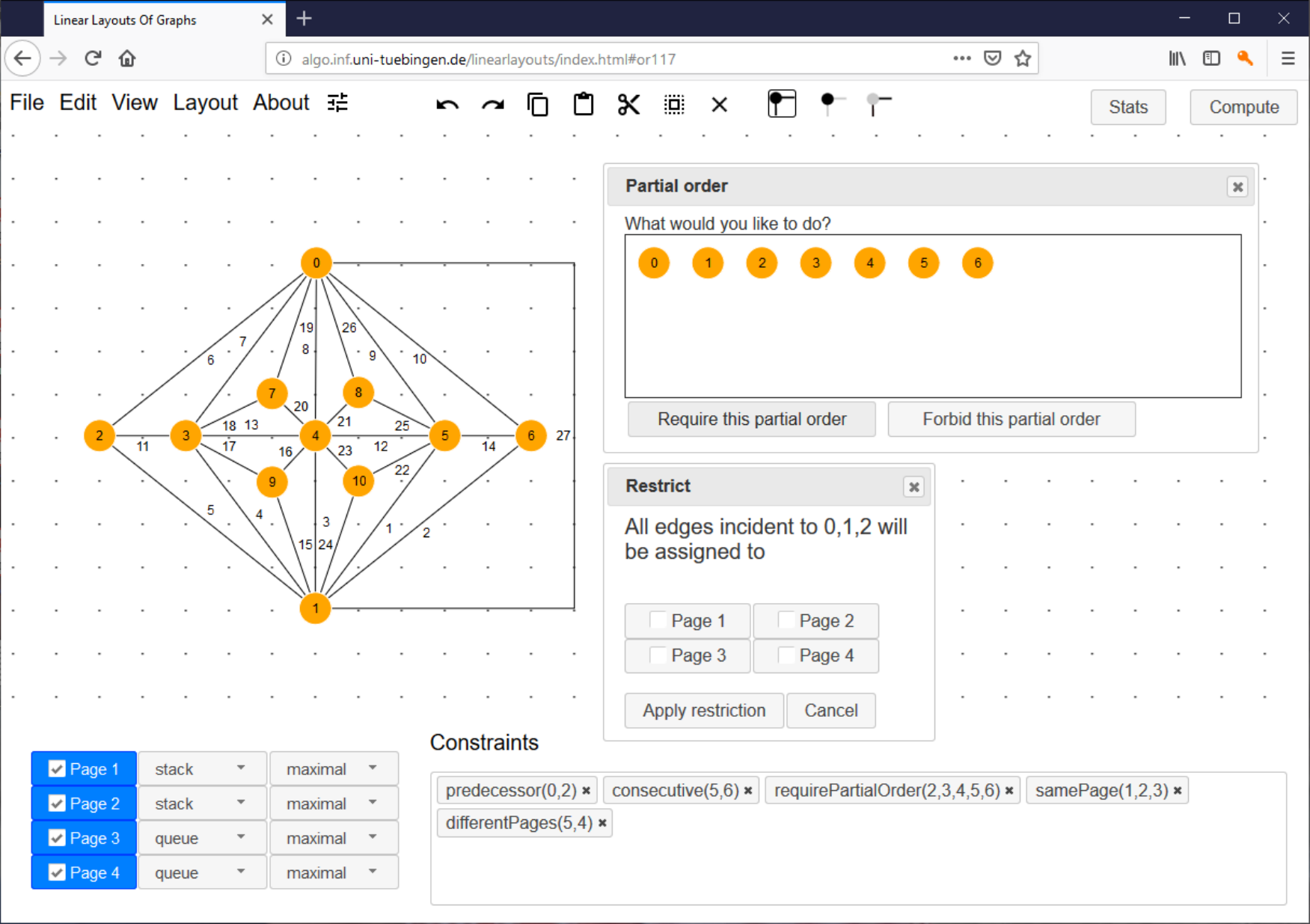}
	\caption{A screenshot of the system in editing mode.}
	\label{fig:editingmode}
\end{figure}

\begin{figure}[t]
	\centering
	\includegraphics[width=\textwidth]{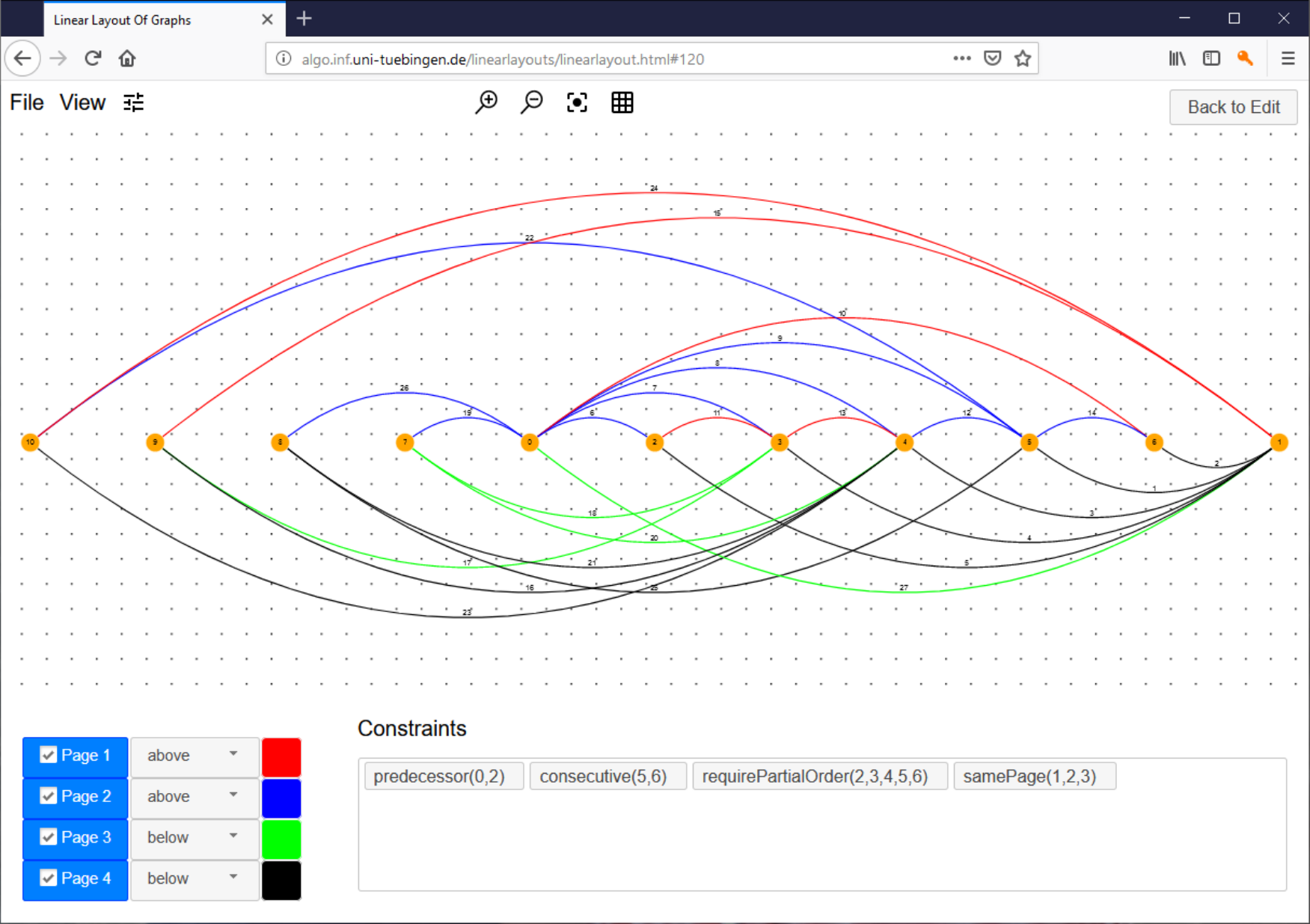}
	\caption{A screenshot of the system in view mode.}
	\label{fig:viewmode}
\end{figure}

The client side is web-based, developed with standard tools (e.g., \texttt{HTML}, \texttt{jQuery} and \texttt{yFiles}~\cite{DBLP:conf/gd/WieseEK00}, which is free for academic usage) and operates in two modes; the \emph{editing} and the \emph{view} (see Figs.~\ref{fig:editingmode} and~\ref{fig:viewmode}).  In the editing mode, the user creates the graph and specifies the constraints of the linear layout to be computed (if any). The graph is created through a graph editor supporting basic interaction with graphs (e.g., creation and deletion of vertices and edges, navigation over the graph, panning, zooming, e.t.c.), that we configured appropriately to meet the needs of a domain expert. The graph editor is accompanied with a \emph{configuration panel} (refer to the bottom part of Fig.~\ref{fig:editingmode}), where the user can configure the type of the layout, as well as, define different constraints on it. More precisely, the following functionalities are currently supported:

\myparagraph{Specification of the type of the linear layout} Through the configuration panel, the user defines the number of available \emph{pages} (i.e., stacks, queues, riques and deques) of the linear layout to be computed. By default, the system supports four pages, in total. However, the total number of pages can be adjusted through the view menu, which allows adding and removing pages. In a subsequent step, the user defines the \emph{type} (i.e., stack, queue, rique or deque) of each of the pages of the linear layout. In this way, the system provides support both for mixed layouts as well; see, e.g,~\cite{DBLP:conf/gd/Pupyrev17}.

\myparagraph{Specification of structural constraints} Besides the number and the type of the available pages of the linear layout, the user can also impose additional structural constraints on the graphs induced by the edges of each of the available pages. Currently, for each of the available pages, the user can choose one of the following options; not to impose any restriction (apart from those that are necessarily imposed by the type of the page), or to restrict the subgraph induced by the edges of a particular page to be either a matching or a tree.

\begin{figure}[t]
	\centering
	\includegraphics[width=0.3\textwidth,page=1]{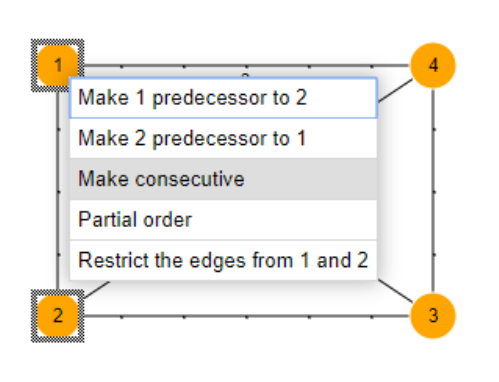}
	\hfil
	\includegraphics[width=0.3\textwidth,page=2]{options}
   \caption{%
   A snapshot illustrating the options available on selected    
   vertices and edges.}
\label{fig:options}
\end{figure}

\myparagraph{Restrictions on the linear order} There exist several ways to impose restrictions on the linear order of vertices of the linear layout (see Fig.~\ref{fig:options}). The constraints are created through the  editor and are stored in a separate component of the configuration panel (so that the user is able~to~remove~them). 

\begin{enumerate}[R.1]
\item \label{r:first-last} \textbf{Specification of first and last vertices}: By selecting a vertex of the graph and by right-clicking on it, the user is able to set it as first or as last in the linear order.
\item \label{r:suc-pre} \textbf{Specification of successors and predecessors of vertices}: By selecting two vertices of the graph and by right-clicking on one of them, the user is able to set one of the selected vertices successor or predecessor of the other.
\item \label{r:consec} \textbf{Specification of vertices to be consecutive}: In the same way, the user is able to require two selected vertices to appear consecutively in the linear order of the vertices. Note that this constraint does not restrict the relative order of them. However, this can be easily achieved by combining Restrictions R.\ref{r:suc-pre} and~R.\ref{r:consec}. 
\item \label{r:order} \textbf{Specification of required and forbidden partial orders.}  By clicking on two or more vertices of the graph, while keeping the ctrl button of the keyboard pressed, the user is able to select multiple vertices in a specific order. Then, by right-clicking on one of them the system provides support to restrict the relative order of the selected vertices to be the one (or not to be the one), in which the vertices were clicked on. 
\end{enumerate}
\noindent\textbf{Restrictions on the edge assignments.} Besides the restrictions on the linear order of the vertices, the user is also able to assign specific edges to particular pages. This can be achieved through the following functions.
\begin{enumerate}[R.1]
\setcounter{enumi}{4}
\item \label{r:edge-page} \textbf{Assignment of edges to the same or to different pages}: By selecting two or more edges of the graph and by right-clicking on one of them, the user is able to instruct the system to assign the selected edges to the same or to different pages of the linear layout. Note that the latter option becomes unavailable, when the number of selected edges is greater than the number of available pages of the linear layout. 
\item \label{r:same-page} \textbf{Assignment of edges to specific pages}: In the same way as above, the user is able to assign selected edges of the graph to one of a set of specific pages of the linear layout. In contrast to Restriction R.\ref{r:edge-page}, the user here has to specify the exact pages, to which the selected pages will be assigned. 
\item \label{r:incident} \textbf{Assignment of edges incident to vertices to specific pages}: The user is also able to assign edges to specific pages through a selection of vertices; in particular, the edges incident to these vertices. In the special case, in which the selected vertices are only two, say $u$ and $v$, then in the linear order of the vertices, $u$ and $v$ define two intervals; the one between $u$ and $v$, and the remaining one. In this particular case, the user is also able to apply the constrain only to the edges from $u$ and $v$ that end to only one of these two intervals. 
\end{enumerate}
\myparagraph{Additional features} The editing mode is equipped with several additional features; in the following, we name few. In order to facilitate the definition of the constraints on the different elements of the graph, the user may restrict the selection mode of the graph editor only to vertices or only to edges, depending on the type of constraints that she wish to introduce. The user is able to save both the graph and its associated constraints in a \texttt{GraphML} file for future considerations (see, e.g., \rurl{graphml.graphdrawing.org}). There exist also two options that are currently supported for exporting the constructed graph; one in \texttt{PDF} and one in \texttt{PNG} format. As side features, the editing mode is also equipped with standard layout algorithms (such as, spring-based, orthogonal and radial), while there is also support for querying the graph for standard properties (e.g., connectivity, acyclicity and planarity). 

Once the creation of the graph and the definition of the constraints on its linear layout have been completed, the user may request to compute the actual linear layout (if any). At this point, the created graph and its constraints are passed to the server side, which is responsible for the actual computation. Once the computation has been completed, the system enters the view mode, where the computed linear layout is presented to the user. In contrast to the editing mode, in the view mode the user can partially interact with the computed linear layout, e.g., 
\begin{inparaenum}[(i)]
\item change the common color and the placement (i.e., above or below the line, on which the vertices reside) of the edges assigned to each page,
\item save or export the final layout to a file, and 
\item navigate, pan or zoom over the layout.
\end{inparaenum}

If the user seeks in further editing the graph and its constraints, then she has to return back to the editing mode. In this transition, the user may choose to work either with the original layout that had constructed before or with the computed linear layout (the user's configurations are also restored). 

\subsection{The Server Side}\label{subsec:server}

The server side of the system has been developed in \texttt{Python}, while for solving the SAT instances, we used the \texttt{Lingeling} solver (\rurl{fmv.jku.at/lingeling}); the source code of the server side is also contained in the \texttt{github} repository mentioned earlier. All requests that arrive to the server are stored in a database (\texttt{SQLite}), which is responsible for associating each of them with a unique id. Once the processing of a request is finished, the computed layout is stored to the database, so to be available to the client at any time. Note that the server side becomes available to multiple clients through a well-documented API available at the server's interface (\rurl{alice.informatik.uni-tuebingen.de:5555}). In other words, any client, that complies with the developed interface, may communicate with the server side and request a linear layout of a given graph under a set of additional constraints that are currently supported at the server. 

The algorithmic core of the server side is an extension of the SAT formulation~\cite{DBLP:conf/gd/Bekos0Z15} mentioned in the introduction, featured with additional functions to overcome known limitations; in particular to support  different types of linear layouts and different types of user-specific constraints. Even though SAT formulations are of limited applicability, and therefore not so common, in graph drawing (with few notable exceptions; e.g.,~\cite{DBLP:conf/gd/BiedlBNNPR13,DBLP:conf/gd/ChimaniZ12,DBLP:conf/gd/GangeSM10}), in this particular scenario the formulation is robust enough to solve non-trivial instances in reasonable amount of time (and, as expected, its performance increases when additional constraints are imposed). In the remainder of this section, we give a short overview of the original formulation followed by a high-level description of the extensions that we made. 

The original formulation makes use of three types of variables $\sigma$, $\phi$ and~$\chi$~with the following meanings: 
\begin{inparaenum}[(i)]
\item for a pair of vertices $u$ and $v$, variable $\sigma(u,v)$ is $\texttt{true}$ if and only if $u$ precedes $v$ in the linear order, 
\item for an edge $e$ and a page $\rho$, variable $\phi_\rho(e)$ is $\texttt{true}$ if and only if edge $e$ is assigned to page $\rho$ of the layout, and 
\item for a pair of edges $e$ and $e'$, variable $\chi(e,e')$ is $\texttt{true}$ if and only if $e$ and $e'$ are assigned to the same page.
\end{inparaenum}  
So, there exist $O(n^2+m^2+pm)$ variables, where $n$ denotes the number of vertices of the graph, $m$ its number of edges, and $p$ the number of available pages. A set of $O(n^3)$ clauses ensure that the underlying order is linear; for  details, refer to~\cite{DBLP:conf/gd/Bekos0Z15}. 
To support different types of layouts, each page $\rho$ is associated with a \emph{type} $\tau(\rho)\in\{\text{\texttt{stack}, \texttt{queue}, \texttt{rique}, \texttt{deque}}\}$ and a different set of at least $\Omega(m^2)$ clauses guarantee the absence of edges in page $\rho$ forming a forbidden pattern in the underlying linear order. Finally, to ensure Restrictions~R.\ref{r:first-last}--R.\ref{r:incident} additional constraints are introduced, e.g., to ensure that a vertex $u$ is a predecessor of a vertex $v$ (see R.\ref{r:suc-pre}), the following clause is introduced: $\sigma(u,v)$.

\section{Proof of Concept}
\label{sec:proof-of-concept}

The system has so far been successful; up to the point of writing, more than 2.000 different linear layouts have been computed using the online version of it. Notably, the system has also been useful in  obtaining results that are of theoretical interest; a summary is given below.  

 \begin{itemize}
\item Using the system, we provided in \cite{DBLP:journals/jocg/KaufmannBKPRU20} a fairly small planar graph, which does not admit a 3-stack layout. This is the smallest known planar graph that requires four stacks. A key to our approach was the introduction of several symmetry-breaking constraints in the SAT instance. These constraints helped to reduce the search space of possible satisfying assignments and made the instance verifiable. 
\item In~\cite{DBLP:journals/tcs/AlamBDGKP21}, we demonstrated concrete
counterexamples to a conjecture by Bernhart and Kainen~\cite{DBLP:journals/jct/BernhartK79} stating that every $\Delta$-regular bipartite graph is dispersable, that is, it admits a $\Delta$-stack layout in which the edges of each page form a matching. These counterexamples form concrete certificates to a corresponding existential proof, which is purely combinatorial. 
\item The system was also very useful in establishing the best-known lower bound of $4$ on the queue number of planar graphs~\cite{DBLP:journals/algorithmica/AlamBGKP20} and in proving that a conjecture by Heath and Rosenberg (asserting that every planar graph admits a 1-stack 1-queue layout~\cite{DBLP:journals/siamcomp/HeathR92}) does not hold even for 2-trees~\cite{DBLP:journals/tcs/AngeliniBKM22}. 
 \end{itemize}

\section{Conclusions}
\label{sec:conclusions}

In this draft, we presented a novel system equipped with several features that automate most of the standard procedures that a domain expert needs for computing different types of (constrained) linear layouts of graphs. We believe that the developed system might be extremely useful for proving lower bounds (e.g., to narrow the current wide gap between the upper and the lower bound on the queue number of planar graphs), which is the main reason to have it available online. We further plan to equip it with additional features, such as adding support for searching and filtering the layouts, support for local and union settings~\cite{DBLP:conf/gd/MerkerU19}, highlighting the restrictions and more advanced definitions of restrictions.

\bibliographystyle{abbrvurl}
\bibliography{stacks,queues,bekos,general}

\end{document}